\let\includefigures=\iffalse
%
\let\useblackboard=\iftrue
%
%
\newfam\black
\input harvmac
\includefigures
\message{If you do not have epsf.tex (to include figures),}
\message{change the option at the top of the tex file.}
\input epsf
\def\figin{\epsfcheck\figin}\def\figins{\epsfcheck\figins}
\def\epsfcheck{\ifx\epsfbox\UnDeFiNeD
\message{(NO epsf.tex, FIGURES WILL BE IGNORED)}
\gdef\figin##1{\vskip2in}\gdef\figins##1{\hskip.5in}
\else\message{(FIGURES WILL BE INCLUDED)}%
\gdef\figin##1{##1}\gdef\figins##1{##1}\fi}
\def\DefWarn#1{}
\def\figinsert{\goodbreak\midinsert}
\def\ifig#1#2#3{\DefWarn#1\xdef#1{fig.~\the\figno}
\writedef{#1\leftbracket fig.\noexpand~\the\figno}%
\figinsert\figin{\centerline{#3}}\medskip\centerline{\vbox{\baselineskip12pt
\advance\hsize by -1truein\noindent\footnotefont{\bf Fig.~\the\figno:} #2}}
\bigskip\endinsert\global\advance\figno by1}
\else
\def\ifig#1#2#3{\xdef#1{fig.~\the\figno}
\writedef{#1\leftbracket fig.\noexpand~\the\figno}%
\global\advance\figno by1}
\fi
\useblackboard
\message{If you do not have msbm (blackboard bold) fonts,}
\message{change the option at the top of the tex file.}
\font\blackboard=msbm10 scaled \magstep1
\font\blackboards=msbm7
\font\blackboardss=msbm5
\textfont\black=\blackboard
\scriptfont\black=\blackboards
\scriptscriptfont\black=\blackboardss

\else

\fi
%
\def\yboxit#1#2{\vbox{\hrule height #1 \hbox{\vrule width #1
\vbox{#2}\vrule width #1 }\hrule height #1 }}
\def\fillbox#1{\hbox to #1{\vbox to #1{\vfil}\hfil}}
\def\ybox{{\lower 1.3pt \yboxit{0.4pt}{\fillbox{8pt}}\hskip-0.2pt}}

\def\comments#1{}

\def\Tr{{{\rm Tr~ }}}

\def\CN{{\cal N}}

\def\II{\relax{I\kern-.10em I}}
\def\IIa{{\II}a}

\def\IZ{\relax\ifmmode\mathchoice
{\hbox{\cmss Z\kern-.4em Z}}{\hbox{\cmss Z\kern-.4em Z}}
{\lower.9pt\hbox{\cmsss Z\kern-.4em Z}}
{\lower1.2pt\hbox{\cmsss Z\kern-.4em Z}}\else{\cmss Z\kern-.4em
Z}\fi}
\def\IB{\relax{\rm I\kern-.18em B}}
\def\IC{{\relax\hbox{$\inbar\kern-.3em{\rm C}$}}}
\def\ID{\relax{\rm I\kern-.18em D}}
\def\IE{\relax{\rm I\kern-.18em E}}
\def\IF{\relax{\rm I\kern-.18em F}}
\def\IG{\relax\hbox{$\inbar\kern-.3em{\rm G}$}}
\def\IGa{\relax\hbox{${\rm I}\kern-.18em\Gamma$}}
\def\IH{\relax{\rm I\kern-.18em H}}
\def\II{\relax{\rm I\kern-.18em I}}
\def\IK{\relax{\rm I\kern-.18em K}}
\def\IP{\relax{\rm I\kern-.18em P}}

%

\def\inbar{\,\vrule height1.5ex width.4pt depth0pt}

\font\cmss=cmss10 \font\cmsss=cmss10 at 7pt
\def\IR{\relax{\rm I\kern-.18em R}}

\def\BZ{\IZ}
\def\BP{\IP}

\def\BC{\IC}

\def\lp10{l_P^{10}}
\def\lp11{l_P^{11}}
\def\R11{R_{11}}

\lref\vafa{C. Vafa, {\it Modular invariance and discrete torsion on
orbifolds}, Nucl. Phys. {\bf B273} (1986) 592.}

\lref\vawit{C. Vafa and E. Witten, {\it On orbifolds with discrete
torsion}, J. Geom. Phys. {\bf 15} (1995) 189.}

\lref\mike{M. R. Douglas, {\it D-branes and discrete torsion}, hep-th/9807235.}

\lref\zaslow{E. Zaslow, {\it Topological Orbifold Models and Quantum 
Cohomology Rings}, Commun. Math. Phys. {\bf 156} (1993) 301.}

\lref\paul{P. Aspinwall {\it Resolution of Orbifold Singularities in String 
Theory}, hep-th/9403123.}

\lref\stable{P. Aspinwall and D. Morrison, {\it Stable singularities in
string theory}, Commun. Math. Phys. {\bf 178} (1996) 115.}

\lref\quivers{M. R. Douglas and G. Moore, {\it D-branes, Quivers, and
ALE instantons}, hep-th/9603167.}

\lref\karpi{G. Karpilovsky {\it Projective representations of finite
groups}, M. Dekker 1985.}

\lref\frac{D.-E. Diaconescu, M. R. Douglas and J. Gomis, {\it Fractional
Branes and Wrapped Branes}, J.High Energy Phys. {\bf 02} (1998) 013.}

\lref\greene {B. R. Greene {\it D-brane topology changing transitions},
Nucl. Phys. {\bf B525} (1998) 284.}

\lref\ooguri{H. Ooguri, Y. Oz and Z. Yin {\it D-branes on Calabi-Yau
spaces and their mirrors}, Nucl. Phys. {\bf B477} (1996) 407.}

\lref\luta{M. Luty and W. Taylor, {\it Varieties of vacua in classical 
supersymmetric gauge theories}, Phys. Rev. {\bf D53} (1996) 3399.}

\lref\corri {E. Corrigan and D.Olive, {\it Colour and Magnetic Monopoles},
Nucl. Phys. {\bf B110} (1976) 237.}

\lref\dak {E. Witten, {\it D-branes and K-theory}, hep-th/9810188.}

\lref\pol{J. Polchinski, {\it Dirichlet Branes and Ramond-Ramond
Charges}, Phys. Rev. Lett. {\bf 75} (1995) 4724.}

\lref\berkdoug{M. Berkooz and M. R. Douglas, {\it Five-branes in
M(atrix) Theory}, Phys. Lett. B395 (1997) 196, hep-th/9610236.}

\Title{\vbox{\baselineskip12pt\hbox{hep-th/9903031}
\hbox{RU-99-11}}}
{\vbox{
\centerline{D-branes and Discrete Torsion II} }}
\centerline{Michael R. Douglas$^{1,2}$ {\it and} Bartomeu Fiol$^{1}$}
\medskip
\centerline{$^1$Department of Physics and Astronomy}
\centerline{Rutgers University }
\centerline{Piscataway, NJ 08855--0849}
\medskip
\centerline{$^2$ I.H.E.S., Le Bois-Marie, Bures-sur-Yvette, 91440 France}
\medskip
\centerline{\tt mrd@physics.rutgers.edu, fiol@physics.rutgers.edu}
\bigskip
We derive D-brane gauge theories for $\BC^3/\BZ_n\times\BZ_n$ orbifolds
with discrete torsion and study the moduli space of a D-brane at a
point.
We show that, as suggested in previous work,
closed string moduli do not fully resolve the singularity,
but the resulting space -- containing
$n-1$ conifold singularities -- is somewhat surprising.
Fractional branes also have unusual properties.

We also define an index which is the CFT analog of the intersection form
in geometric compactification, and use this to show that the
elementary D$6$-brane
wrapped about $T^6/\BZ_n\times\BZ_n$ must have $U(n)$ world-volume gauge
symmetry.

\Date{March 1999}
\newsec {Introduction}
String theory can be well defined
on spaces that from the classical geometric point of view have
singularities.  In recent years, D-branes have brought a new perspective
on this.

In this work we will study D-branes at a $\BC^3/\BZ_n\times \BZ_n$
singularity with discrete torsion, following the ideas of
\refs{\vafa,\vawit,\mike}.
Discrete torsion
was defined in {\vafa} for orbifolds $M/\Gamma$ in world-sheet terms:
different sectors in the closed string path integral distinguished by
their twisted boundary conditions are weighed by phases.
The first example is genus one with two twists $g$ and $h$ corresponding
to the two generators of $\pi_1$ and a weight $\epsilon(g,h)^2$. Higher
loop modular invariance requires $\epsilon (g,h)$ to be a cocycle
in $H^2 (\Gamma, U(1))$.  For $\Gamma= \BZ_n\times \BZ_n$ this is $\BZ_n$
and we study the case where $\epsilon (g,h)$ is a generator of this group
(``minimal discrete torsion'').

Singularities can appear as degenerations in Calabi-Yau manifolds
and from the point of view of algebraic geometry admit two types of
desingularization.
Singularities caused by degeneration of the complex
structure can be deformed, while singularities caused by degeneration of
the K\"ahler form can be resolved (or blown up).
As part of a compact space the possible resolutions depend on the global
topology of the space: harmonic $(2,1)$-forms correspond to complex structure
deformations, and harmonic $(1,1)$-forms to K\"ahler form deformations.
The analysis of the conformal field theory of strings on
$\BC^3/\BZ_n\times \BZ_n$ uses standard orbifold techniques and
we review these results.
While string theory without discrete torsion allows only
K\"ahler deformations, the theory with minimal discrete torsion allows
only complex structure deformations.   In the latter case,
although conformal field theory does not lead to a clear geometric
picture of these deformations,
one can make hypotheses based on parameter counting and
other general considerations which strongly suggest that the complex
structure deformations provided by the string theory do {\it not} allow
completely resolving the singularity.  In {\vawit} it was suggested
that the $n=2$ model contains an irresolvable conifold singularity,
while the $n=3$ model could contain an irresolvable singularity
of codimension $2$.

By studying a D-brane in this background, one sees space-time arise
in a different way. The low energy motion of D-branes is described by a
gauge theory, and space-time appears as its
moduli space. The two types of deformation are then obtained
as modifications of the F-flatness conditions (complex structure) and
of the D-flatness conditions (K\"ahler form).

World-volume theories for D-branes at orbifold singularities can be obtained
as projections of $\CN=4$ super Yang-Mills theory and
in {\mike} it was argued that discrete torsion is implemented by doing
this with projective representations of $\Gamma$,
and the $\BC ^3/\BZ_2\times \BZ_2$ case was studied.
The resulting moduli space was exactly as predicted in
{\vawit}: after turning on all moduli, an isolated singularity remains.

In the present paper we extend this to minimal discrete torsion for
$\BC^3/\BZ_n\times\BZ_n$ and arbitrary $n$. We find that for arbitrary $n$,
after turning on all the available complex structure deformations, the three
fixed lines are resolved, but $n-1$ conifold singularities appear.

In section 2 we present the
orbifolds and find the closed string marginal operators
for various choices of discrete torsion. In section 3 we
derive the gauge theory, in section 4 describe the generic
branch of moduli space, and
in section 5 study special branches which appear at partial resulutions.

The interpretation of fractional branes is discussed in section $6$.
There are many differences from the case without discrete torsion;
some puzzles are listed, including one on charge quantization.
We point out that
charge quantization is best formulated in terms of an index,
${\rm Tr}_{ab}\ (-1)^F$
in the open string sector with boundary conditions $a$ and $b$, and resolve
the puzzle in this case.
Section 7 summarizes the conclusions.

\newsec {$\BC ^3/\BZ_n \times \BZ_n$ orbifolds and discrete torsion}

In this section we introduce the orbifolds we are going to study and discuss
their possible resolutions. Let $\BC ^3$ be described by three complex
coordinates, $z_1, z_2, z_3$. The group $\Gamma = \BZ_n\times \BZ_n$
is generated by two elements $g_1$ and $g_2$.
We will denote a generic element $g_1^ag_2^b$ of
$\Gamma $ as $(a,b)$.

As an action of $\BZ_n \times \BZ_n$ on $\BC ^3$, we take $R(g)$ defined by
\eqn\staction{\eqalign{
 & g_1:(z_1,z_2,z_3) \rightarrow (z_1,e^{-{2\pi i\over
n}}z_2, e^{2\pi i\over n} z_3)\cr & g_2:(z_1,z_2,z_3) \rightarrow (e^{2\pi i
\over n} z_1,z_2,e^{-{2\pi i\over n}} z_3).
}}
This is the unique faithful representation in the sense that
any $R'(g)=R(H(g))$ for some group homomorphism $H$.

The 2-cocycle classes of $H^2(\Gamma, U(1))\cong \BZ_n$ are represented by
$$\matrix {& \epsilon ^m(g,h) :& \Gamma \;\; \times \;\; \Gamma &
\rightarrow & U(1)\cr & &\left ((a,b),(a',b')\right )&\rightarrow &\zeta ^{m
(ab'-a'b)}}$$ where $\zeta =e^{\pi i\over n}$ for $n$ even and
$\zeta =e^{2\pi i\over n}$ for $n$ odd. The different values of $m=0...n-1$
correspond to the different elements of $H^2(\Gamma, U(1))\cong
\BZ_n$. In this paper we will study the theories that arise when
$(m,n)=1$, that is, when $\zeta ^{2m}$ is a generator of $\BZ_n$.

\subsec {Possible resolutions of $\BC ^3/\BZ_n\times \BZ_n$}

One can define $\BC ^3/\BZ_n\times \BZ_n$ as an affine variety in $\BC^4$,
given by $F(x,y,z,t)=0$, where $F(x,y,z,t)=xyz-t^n$. The $\Gamma $-invariant
variables are $m_i=(z_i)^n$ and $b=z_1z_2z_3$, related by $m_1m_2m_3=b^n$.
This variety presents non-isolated singularities, in the form of three
(complex) fixed lines of $\BC ^2/\BZ_n$ singularities: $(z_1,0,0)$,
$(0,z_2,0)$ and $(0,0,z_3)$. These three lines intersect at the
origin, that is a fixed point under the action of the whole group $\Gamma $.

Using toric geometry it is fairly straightforward to give a
description of the possible blow-up resolutions of $\BC ^3/\BZ_n\times
\BZ_n$. The fact that the singularities are non-isolated does not
affect the discussion at this level. The discussion parallels the one for
$\BC^2/\BZ_n$ \paul, so we will be brief. The fan for $\BC ^3/\BZ_n \times
\BZ_n$ has a single big cone, determined by three vectors $(0,0,1)$,
$(n,0,1)$ and $(0,n,1)$. The first intersection (away from the origin) of
the three edges of the cone with the lattice determines the hyperplane $z=1$,
so the singularities are Gorenstein. The intersection is a triangle in the
hyperplane, with vertices $(0,0,1)$, $(n,0,1)$ and $(0,n,1)$. The resolution
is far from unique: the possible blow-up resolutions are represented by
different triangulations of this triangle. For a given $n$ there are many
possible triangulations, all related by flop transitions.

In \greene, the case $n=2$ without discrete torsion was studied using
D-brane methods.  As we discuss below the moduli are K\"ahler moduli
and it was shown using toric methods that all the possible resolutions
can be obtained from the D-brane gauge theory.


Another way to resolve the orbifold is to deform its complex
structure. To find the possible relevant deformations, first we
compute the ideal generated by the partial derivatives of $F$:
$J=\{\partial _iF\}=\{yz,xz,xy,t^{n-1}\}$. The quotient of the ring of
polynomials in $x,y,z,t$ by $J$ gives us the set of relevant
deformations of $F$:

\eqn\quot{Q={C[x,y,z,t]\over \{\partial _iF\}}=\{x^a,y^b,z^c,1,t..t^{n-2}\}}

The quotient ring is infinite dimensional, something that matches with the
fact that we have non-isolated singularities. If we deform $F$ adding a
power of $x$, we resolve a complex line (and similarly for $y$ and $z$). If
we deform by $1$ or $t$, we resolve completely the variety. If we resolve
by $t^r$ with $1<r\leq n-2$ we reduce the singularities to those of $xyz-t^r$.

We see that mathematically these orbifolds can be completely
resolved by deformation of their complex structure.
This does not necessarily mean that all of these deformations
are available in the physical theory -- at the very least, each physical
deformation (modulus) must correspond to a marginal operator in
closed string theory on the orbifold.  Thus we need to compute this spectrum.

\subsec {Cohomology}

Space-time supersymmetry relates marginal operators to
ground states in the Ramond sector.
As is well known, for sigma model compactification these are in
one-to-one correspondence with the cohomology of the target space,
so if the orbifold is the limit of a smooth target space $M$ this computation
will give us $H^*(M)$.  We will also think of the results as defining
the cohomology of the orbifold.
As we are about to see, it
depends markedly on whether we introduce discrete torsion or not.

We follow the procedure used in \vawit\ (see also \zaslow).
There are $n^2$ sectors in the theory: one untwisted sector corresponding
to the element $(0,0)$ and $n^2-1$ twisted sectors corresponding to the
remaining elements of $\Gamma $.
For each sector $g$, we let $M_g$ be the subset of
$\BC ^3$ fixed under the action of $g$.
In the ordinary case, without discrete torsion, the usual sigma model
analysis generalizes to show that the ground states in the $g$-twisted
sector correspond to forms $H^{p,q}(M_g)$
that are invariant under the action of $\Gamma$:
\eqn\forms{
R(h)w=w \qquad \forall h\in \Gamma .
}
Such a form $\omega^{p,q}$ will contribute to
$H^{p+s,q+s}(M)$ with $s$ a computable function of $g$:
for $g$
acting as $z_i \rightarrow e^{i\theta _i} z_i$ with
$0\leq \theta _i <2\pi$, it is given by
\eqn\shift{
s=\sum _i {\theta _i\over 2\pi}.
}
This shift is derived for example in {\zaslow} as the shift
in the fermion number and $U(1)$ charges
of the vacuum for the twisted sector $g$.

Let us start with the untwisted sector $(0,0)$. The fixed point set in this
case is of course $\BC ^3$ itself, but not all the forms defined on $\BC ^3$
are
$\Gamma$-invariant. For instance $dz_1\wedge d\bar z_1$ is invariant and it is
kept, but $dz_1\wedge dz_2$ is not invariant and it is projected out. The
contribution to the Hodge diamond coming from the untwisted sector is, for
$n>2$

$$\matrix {& & & &1& & &  \cr
           & & &0& &0& &  \cr
           & &0& &3& &0&  \cr
           &1& &0& &0& &1 \cr
           & &0& &3& &0&  \cr
           & & &0& &0& &  \cr
           & & & &1& & &   }$$

For $n=2$ the untwisted sector has instead $h^{2,1}=h^{1,2}=3$, since
for instance the form $dz_1 \wedge dz_2 \wedge d\bar z_3$ survives.

Now let's consider the $n^2-1$ twisted sectors. We are going to discuss the
twisted sector corresponding to the element $(a,b)$. First we
have to look for the fixed point set of the element $(a,b)$. There are two
classes of elements, according to their fixed point sets. The $3(n-1)$
elements of the form $(a,0)$, $(0,b)$ and $(a,a)$ have a (complex) line of
fixed points: $(z_1,0,0)$, $(0,z_2,0)$ and $(0,0,z_3)$ respectively, as can be
seen from \staction. The remaining $(n-1)(n-2)$ elements have only the origin
as their fixed point set.

As a warm-up, we compute the contribution of the twisted sectors for
the theories without discrete torsion. For each of the $3(n-1)$ elements of
the kind $(a,0)$, $(0,b)$ and $(a,a)$,
the cohomology before projecting is generated by $1,dz, d\bar z$ and
$dz\wedge d\bar z$. Without discrete torsion $1$ and $dz\wedge d\bar {z}$ are
$\Gamma $-invariant, whereas $dz$ and $d\bar {z}$ are not.
The shift for these $3(n-1)$ elements is easily computed to be $s=1$ for all
of them, so in the case of theories without discrete torsion, the contribution
of these $3(n-1)$ twisted sectors to the Hodge diamond is $h^{1,1}=h^{2,2}=3
(n-1)$. For the remaining $(n-1)(n-2)$ elements, the origin is the only fixed
point,
and in absence of discrete torsion, the $(0,0)$-form $1$ is $\Gamma$-invariant.
After the shift, half of these elements contribute to $H^{1,1}$, and the other
 half contribute to $H^{2,2}$. All in all, in absence of discrete torsion, the
 contribution of the twisted sectors to the Hodge diamond is

$$\matrix {& & & &0& & &  \cr
           & & &0& &0& &  \cr
           & &0& &{(n+4)(n-1)\over 2}& &0&  \cr
           &0& &0& &0& &0 \cr
           & &0& &{(n+4)(n-1)\over 2}& &0&  \cr
           & & &0& &0& &  \cr
           & & & &0& & &   }$$

After this warm-up, let's turn to the theories with discrete torsion. The
whole discussion depends on $r=\gcd(m,n)$, and the results differ markedly
between the cases $r=1$ ($m$ and $n$ are relative primes) and $r>1$.
Now in the sector $g$ we keep the forms $w$ such that

\eqn\keepw{
\left (\epsilon (g,h)\right )^2\; R(h)w=w\qquad \forall h\in \Gamma
}

Since $\epsilon (1,h)=1 \; \forall h$, the untwisted sector remains the same,
so we go to the twisted sectors.
Let's deal first with the $3(n-1)$ sectors corresponding to elements
that fix a complex line. Recall that before projecting the cohomology
is given by $1,dz,d\bar z$ and $dz\wedge d\bar z$. For $1$ and $dz\wedge
d\bar z$, $R(h)w=w$, so they survive if $ \left (\epsilon (g,h)\right )^2=1$,
and this happens for $3(r-1)$ elements of these sectors. Furthermore, if $r=1$
there are exactly 3 elements for which $dz$ is kept and 3 elements for which
$d\bar z$ is kept. If $r>1$ then $dz$ and $d\bar z$ are projected out. As
before, the shift for these elements is 1, but now the forms that
we keep are in $H^{0,1}$ and $H^{1,0}$ of the respective fixed point sets, so
they contribute to $H^{1,2}$ and $H^{2,1}$ of the whole orbifold.
Finally, we have to consider the remaining twisted sectors, associated to
elements that only leave fixed the origin. Now the only element in the
cohomology is $1$, and it is kept by $(r-1)(r-2)$ elements, half of them
contributing after the shift to $h^{1,1}$ and half of them to $h^{2,2}$.

Putting together these facts, the twisted sector contribution to the Hodge
diamond for the $r=1$ case is:

$$\matrix {& & & &0& & &  \cr
           & & &0& &0& &  \cr
           & &0& &0& &0&  \cr
           &0& &3& &3& &0 \cr
           & &0& &0& &0&  \cr
           & & &0& &0& &  \cr
           & & & &0& & &   }$$

and for $r>1$

$$\matrix {& & & &0& & &  \cr
           & & &0& &0& &  \cr
           & &0& &{(r+4)(r-1)\over 2}& &0&  \cr
           &0& &0& &0& &0 \cr
           & &0& &{(r+4)(r-1)\over 2}& &0&  \cr
           & & &0& &0& &  \cr
           & & & &0& & &   }$$

When $r>1$ the effect of discrete torsion is just to decrease the number of
ground states in $h^{1,1}$ and $h^{2,2}$ compared to the case without
discrete torsion (that can be considered a particular case, $m=0$, $r=n$). On
the other hand,
when $r=1$ there are no K\"ahler deformations available, and we find
the same number of complex deformations for every $n$. Thus the general
$\BZ_n\times \BZ_n$ singularity will have either complex structure or
K\"ahler deformations, but not both.

\newsec {The D-brane gauge theories}

In this section we derive the gauge theory that describes D-branes on the
$\BC ^3/\BZ_n \times \BZ_n$ orbifold with discrete torsion for
$\gcd(m,n)=1$.
After presenting the gauge theory, we look for its moduli space and check
that, for the regular
representation, it reproduces the original orbifold. We then include
the modification of the superpotential due to the 3 possible complex
structure deformations that we found in the previous section, and
determine the corresponding deformed moduli space. As we will see, the
addition of the complex structure deformations resolves the fixed lines of the
original orbifold, leaving $n-1$ conifold singularities.

\subsec{Derivation of the quotient theory}

The general procedure
is by now well known \quivers. There are two choices that we have to
make: the action of $\Gamma$ on space-time, $R(g)$, and the action of
$\Gamma $ on the Chan-Paton indices, $\gamma (g)$. If we want to
describe $N$ independent D-branes on the orbifold, we start by placing
$|\Gamma | N$ D-branes on the original $\BC ^3$. Initially this
corresponds to a $U(|\Gamma |N)$ SYM theory with ${\cal N}=4$ in $d=4$. To
construct the quotient gauge theory we impose projection conditions for the
gauge and the scalar fields
\eqn\projcon{\eqalign{
& A=\gamma(g)A\gamma(g)^{-1}\cr & Z^i=R^{ij}(g)\left (
\gamma(g)Z^j\gamma(g)^{-1}\right )
}}
Plugging back the fields surviving these projections into the original
$U(|\Gamma |N)$ theory, we obtain the quotient gauge
theory.

Our choice of the space-time action is \staction; since this
lies in $SU(3)$, the corresponding theory will have
${\cal N}=1$ in $d=4$. Where does the discrete torsion enter the game?
As observed in \mike, it is implemented in the gauge
theory by taking the action on the Chan-Paton indices to be a
{\it projective} representation of $\Gamma $, that is $\gamma (g)\gamma
(h)=\epsilon (g,h)\gamma (gh)$.

According to {\karpi}, when $\epsilon \equiv \zeta ^{2m}$ is a primitive
$n$'th root of $1$ (true when $\gcd(m,n)=1$), $\BZ_n \times \BZ_n $
has a unique irreducible projective representation.
It is
\eqn\rep{
\gamma_1 (g_1)= P\qquad \qquad \gamma_1 (g_2)=Q
}
where the matrices $P$ and $Q$ are as follows.
For $n$ odd,
\eqn\matrius{
P=\pmatrix {&0 &1 &0 &\dots &0 \cr
                        &0 &0 &1 &\dots &0 \cr
                       &\dots &\dots &\dots &\dots &\dots \cr
                        &0 &0  &\dots &0 &1 \cr
                        &1 &0  &0  &\dots &0 }\qquad
Q=\pmatrix {&0 &\epsilon &0 &\dots &0 \cr
                        &0 &0 &\epsilon ^2 &\dots &0 \cr
                       &\dots &\dots &\dots &\dots &\dots \cr
                        &0 &0  &\dots &0 &\epsilon ^{n-1} \cr
                        &1 &0  &0  &\dots &0.
}}
For $n$ even, $P$ is as above, define $\delta ^2=\epsilon$
and
\eqn\matriuss{
Q=\pmatrix {&0 &\delta &0 &\dots &0 \cr
                        &0 &0 &\delta ^3 &\dots &0 \cr
                       &\dots &\dots &\dots &\dots &\dots \cr
                        &0 &0  &\dots &0 &\delta ^{2n-3} \cr
                        &\delta ^{2n-1} &0  &0  &\dots &0
}}

Note that $PQ=\epsilon QP$, so although $\Gamma $ is abelian,
the projective representation is not. The general representation is
a direct sum of $M$ copies of this:
$$\gamma_M = \gamma_1 \otimes {\bf 1}_M.$$
In particular, the regular representation of dimension $n^2$ is $\gamma_n$.

We can now give the solution of the projection \projcon\ for
every $m$ such that $(m,n)=1$.
By composing the space-time action
\staction\ with a $\BZ_n\times\BZ_n$, the general solution
based on the irreducible representation \rep\ can be brought to the form
\eqn\solproj{A=\hbox {I}\qquad Z_1 = P\qquad Z_2 = Q\qquad Z_3 = (PQ)^{-1}.
}
The most general solution is obtained by tensoring this $n\times n$
solution with $M\times M$ matrices.

Intuitively we expect a theory describing $N$
'true' branes away from the origin will involve $|\Gamma |N=n^2N$ images and
thus use $N$ copies of the regular representation, i.e.
$\gamma_M$ with $M=Nn$.\footnote {$^1$}{
There is a simple argument why one expects to need $M\ge n$ to
get a three complex dimensional moduli space.  One can see that supersymmetric
vacua (in the undeformed orbifold theory) are still described by
{\it commuting} matrices in the original $U(Mn)$ theory; to make
commuting matrices out of the solution \solproj\ one must tensor them
with another projective representation with the opposite cocycle,
which will have dimension $M\ge n$.}
On the other hand string consistency conditions do not require
$M$ to be a multiple of $n$; we
will return to the interpretation of this possibility later.

After
tensoring \solproj\ with $M\times M$ matrices and substituting into the
the $U(Mn)$ SYM theory, we obtain the quotient theory, which we proceed
to describe.

\subsec{The orbifold theory}

The gauge theory describing $\BC^3/\BZ_n\times\BZ_n$ with minimal
discrete torsion is a $U(M)$ theory
with three chiral superfields $\phi_i$, $i=1,2,3$ in the adjoint. The
superpotential is
\eqn\super{W=\hbox { tr}\{\phi _1 (\phi _2 \phi _3 -\epsilon ^{-1}\phi _3
\phi _2)\}}

This leads to the F-flatness conditions
\eqn\fflat{\phi _i \phi _j -\epsilon ^{-1}\phi _j \phi _i=0 \qquad
i\neq j}
and D-flatness conditions
\eqn\dflat{ \sum _i [\phi _i,\phi _i ^\dagger]=0.}

The superpotential \super\ preserves a $U(1)^3$ subgroup of the $SU(4)$
R-symmetry of $\CN=4$ super Yang-Mills, the individual phase rotations
of the $\phi_i$.  The diagonal $U(1)$ is the usual $\CN=1$ R-symmetry.

So far, we haven't introduced in the gauge theory the three complex structure
deformations that we found in the previous section. These appear
as the following deformation of the superpotential of the gauge theory:
\eqn\newsp{\Delta W= \sum _i \zeta _i \hbox { tr } \phi _i.}

This is consistent with the
$U(1)^3$ symmetry if we assign $\zeta _1$ the charges $(0,1,1)$,
and similarly for $\zeta _2$ and $\zeta _3$.
The symmetry will then prohibit corrections higher order in $\phi$.%
\footnote{$^2$}{
It can be checked by world sheet computation that the
disk diagram with a bulk insertion of the twisted closed string
operator $V(\zeta_i)$ and a boundary insertion of $V(F_i)$, the auxiliary
field in the $\chi_i$ multiplet, is non-zero.  The $U(1)^3$ symmetry is the
unbroken subgroup of $SO(6)$ rotations around the fixed point and one can
also check that $V(\zeta _i)$ has the charges stated above.
}

The new term modifies the F-flatness conditions to
\eqn\dfflat{\eqalign{
 & \phi _1 \phi _2-\epsilon ^{-1}\phi _2 \phi _1 = -\zeta _3\cr
 & \phi _2 \phi _3-\epsilon ^{-1}\phi _3 \phi _2 = -\zeta _1\cr
 & \phi _3 \phi _1-\epsilon ^{-1}\phi _1 \phi _3 = -\zeta _2
}}

The D-flatness conditions \dflat\ remain the same. Our goal will be to find
the new moduli space, corresponding to these deformed conditions.

\newsec{Moduli spaces as varieties}

In our discussion of the moduli space, we will use the gauge invariant
polynomials\footnote{$^3$}{Our normalizations
differ from those of \mike.}
\eqn\defpol{
M_{ij\dots k}={1\over n}\hbox { tr }\{\phi _i \phi _j
\dots \phi _k\}
}

Taking the trace of the F-flatness conditions \dfflat\ we find
\eqn \mij{M_{ij}=-\left (1-\epsilon ^{-1}\right )^{-1}\zeta _k
\equiv \xi _k}

Let us first solve the F and D conditions in the undeformed theory. We start
with the regular representation. For $M=n$ (that is, $N=1$, a single D-brane),
recalling that $PQ=\epsilon QP$, we see that a solution is given by

\eqn\sol{
\phi _1 = z_1Q \qquad \phi _2 = z_2 P\qquad \phi _3 =z_3 (QP)^{-1}
}

What is the equation for the moduli space? If we define $n_1$ as the
number of $\phi _1$'s in a given gauge invariant polynomial $M_{ij\dots k}$,
and similarly for $n_2$ and $n_3$, the only non-zero gauge invariant
polynomials have $U(1)^3$ charges $n_1\equiv n_2 \equiv n_3\; (n)$.
In particular

\eqn\solu{\eqalign{
& M_{\underbrace {11\dots 1}_n}=z_1^n \cr
                   & M_{\underbrace{22\dots 2}_n}=z_2^n \cr
                 & M_{\underbrace {33\dots 3}_n}=(-1)^{n-1}z_3^n \cr
                 & M_{123}=z_1 z_2 z_3
}}
so it is natural to identify these gauge invariant polynomials with
the $\Gamma $-invariant variables $m_i$ and $b$ of $\BC^3/\BZ_n \times \BZ _n$,
and conclude that the undeformed moduli space is
\eqn\moduli{
M_{\underbrace {11\dots 1}_n}M_{\underbrace{22\dots 2}_n}
M_{\underbrace {33\dots 3}_n}=(-1)^{n-1}(M_{123})^n.
}

{}From now on, we drop the brackets, and unless otherwise stated, it is
understood that $M_{ii..i}$ has $n$ indices.

What happens if we don't take the regular representation?
For $1\leq M < n$, taking determinants on both sides of the original
F-flatness conditions \fflat\ and recalling that $\epsilon$ is a primitive
n$th$-root of 1, we readily see that $|\phi _i| |\phi _j|=0$, so at least two
out of three matrices have zero determinant. By going to a basis where one
of these matrices with zero determinant, say $\phi _1$, is diagonal, we can
easily argue that most of the off diagonal elements of $\phi _2$ and $\phi _3$
have to vanish in order to satisfy the F-flatness conditions, and then, for
$M<n$, the remaining off diagonal elements have to vanish in order to satisify
the D-flatness conditions. Therefore, for $M<n$, all the solutions of the
F and D flatness conditions are diagonal matrices, and there are no Higgs
branches. For $M=1$, the moduli space is
given by the three lines $M_1=M_2=0$, $M_2=M_3=0$ and $M_1=M_3=0$.

We turn to the deformed case. For the regular representation, there is a
simple three parameter family of solutions of the deformed F-flatness
conditions \dfflat:

\eqn\soldef{
\phi _1 =z_1Q+\xi _3{P^{-1}\over z_2}\;\;\; \phi _2=z_2P+\xi _1
{QP\over z_3}\;\;\;\phi _3=z_3(QP)^{-1}+\xi _2{Q^{-1}\over z_1}
}

These do not satisfy the D-flatness conditions and it appears to be quite
hard to find explicit solutions of both conditions.  On the other hand
we are guaranteed by general results that, for every solution of the
F-flatness conditions, there will exist a unique gauge equivalence class
of solutions of the D-flatness conditions \luta.
In the present case it is defined by minimizing the functional
$f = \Tr \sum_i \phi_i\phi_i^+$ over a gauge orbit
$\phi_i\rightarrow g\phi_i g^{-1}$ with $g\in GL(n)$ the
complexified gauge group.
We do not need to find these solutions explicitly in order
to describe the moduli space with its complex structure;
they would have been useful
if (for example) we wanted to compute the metric.

For many purposes it is more useful to describe the moduli space as
a variety, i.e. as the solutions of a set of polynomial equations between
gauge invariant polynomials.  In particular this will bypass the need to
determine which of the solutions \soldef\ are gauge equivalent.
An overcomplete set of such equations is
obtained by combining all equations $\Tr W'(\phi) P(\phi)=0$ for all
polynomials $P$ with the complete set of identities on gauge-invariant
polynomials constructed from matrices of dimension $n$.  The latter are
quite simple for $n=2$ and this approach was followed in \mike;
however for $n>2$, the system of relations between gauge invariant polynomials
is very complicated.

The observation that makes the problem tractable
is that, if we are satisfied just to study the moduli space
of the gauge theory, then relations valid {\it on the moduli
space} are good enough for us.
In particular, the F-flatness conditions \dfflat\ turn
out to be a powerful constraint that allows us to establish
many relations among the gauge invariant polynomials that, though not valid
in general, hold on the moduli space. In fact, we can argue that on
the moduli space of this $U(n)$ gauge theory, when all $\zeta\ne 0$,
any non-zero gauge
invariant polynomial can be written in terms of the 4 gauge invariant
polynomials used to define the undeformed moduli space! The proof of this
statement is presented in the appendix.

Accepting this result, we conclude that the deformed moduli space is indeed
a variety in $\BC^4$.  As we argued the branch we found is three dimensional
and by general results in algebraic geometry a three dimensional
subvariety must be a hypersurface; i.e.
the solution of a polynomial equation in $\BC^4$.
This equation must respect the $U(1)^3$ symmetry
acting on $\phi_i$ and $\xi_i$ as well as an obvious permutation symmetry.

It is natural to suppose that this equation would be a deformation of
\moduli\ with corrections polynomial in the $\xi$'s.
Adding all polynomial perturbations respecting the symmetries we
arrive at the ansatz
\eqn\guess{M_{11..1} M_{22..2}M_{33..3}+c(M_{11..1}\xi _1^n
+M_{22..2}\xi _2^n+M_{33..3}\xi _3^n)=\sum _{k=0}^{[n/2]}
a_k M_{123}^{n-2k}(\xi _1\xi _2\xi _3)^k}
for the equation determining this branch of moduli space.

We can determine the coefficients by requiring that the solutions \soldef\
are solutions of \guess.  The corresponding invariants are
\eqn\gaupol{\eqalign{& M_{11..1}=z_1^n+{\xi_3^n\over z_2^n}\cr
                 & M_{22..2}=z_2^n+(-1)^{n-1}{\xi_1^n\over z_3^n}\cr
                 & M_{33..3}=(-1)^{n-1}z_3^n+{\xi_2^n\over z_1^n}\cr
& M_{123}=z_1z_2z_3+\epsilon ^{-1}{\xi _1\xi _2\xi _3\over z_1z_2z_3}}}
and plugging these into the proposed relation \guess, we find
\eqn\coeffs{
c=-1;\qquad
a_k=(-1)^{n+k-1}\epsilon ^{-k}{n\over n-k} {n-k \choose k}.
}

Now that we have checked the equation \guess\ on the explicit solutions
\soldef, we can assert that it {\it is} the equation defining this branch of
moduli space; the hypothesis that it might take the form \guess\ has
been explicitly verified.
In general the equations defining the moduli space derived by following
the general procedure could have additional solutions which will be
additional branches of the moduli space.  Let us focus for now on this branch.

The coefficients $a_k$ correspond to the $n$'th Chebyshev polynomial of the
first kind: $T_n (\cos \theta)= \cos n\theta$, and thus the equation
describing this branch of the moduli space takes a simple form:
it is $F(M_{ii..i}, M_{123})=0$ where
\eqn\defmod{\eqalign{
& F(M_{11..1},M_{22..2},M_{33..3},M_{123}) = \cr
&M_{11..1} M_{22..2}M_{33..3}-M_{11..1}\xi _1^n-M_{22..2}\xi _2^n-M_{33..3}
\xi _3^n \cr
&\qquad\qquad\qquad
- 2(-1)^{n-1}\chi ^n T_n\left ({M_{123}\over 2\chi}\right )
}}
where we defined $\chi =({\epsilon ^{-1}\xi_1\xi_2\xi_3})^{1/2}$.

This is not the general deformation of \moduli\ and thus this variety
might generically be singular.
Singularities will be
solutions of $\partial F/\partial M_{ii..i}=\partial F/\partial M_{123}=F=0$.
To start with, $\partial F/\partial M_{ii..i}=0$ requires that
\eqn\fsing{M_{ii..i}= + \left ({\xi _j\xi _k\over \xi _i}\right )^{n/2}}
or
\eqn\ssing{M_{ii..i}= - \left ({\xi _j\xi _k\over \xi _i}\right )^{n/2}}
so at the singularity $F(M_{ii..i},M_{123})=2(\xi _1\xi _2\xi _3)^{n\over 2}
\left (\mp 1-T_n(M_{123}/2\chi)\right )$. We have to determine if any
of the $n-1$ solutions of ${\partial F\over \partial M_{123}}=0$ is also a
solution of $F=0$ for these values of $M_{ii..i}$. Defining $\cos
\theta =(M_{123}/2\chi)$, the $n-1$ roots of
${\partial F\over\partial M_{123}}$ are given by
\eqn\rootheta{\theta _k ={\pi k\over n}\qquad k=1,..,n-1}
and since $T_n(\cos \theta _k)= (-1)^k$,
we see that the variety has $n-1$ singularities:
\eqn\thesing {M_{ii..i}= (-1)^{k-1} \left ({\xi _j\xi _k\over \xi _i}
\right )^{n/2}\;\;\; M_{123}=2\chi cos {\pi k\over n}\qquad k=1,..,n-1}
It is easy to check that these are conifold singularities. To do so, expand
\defmod\ around any of the singularities $M_{ii..i}=M_{ii..i}^r+x_i$ and
$M_{123}=M_{123}^r+t$, where $M_{ii..i}^r$ and $M_{123}^r$ are the values
at the $r$'th singularity \thesing . The result is
\eqn\coni{\left ({\chi ^{n-2} (-1)^{r-1} n^2\over 4\sin ^2 \theta_r}
\right ) t^2+{\cal O}(t^3)=\sum _{i<j} M_{kk..k}^r \;x_ix_j+{\cal O}(x^3)}
and since the determinant of this quadratic form is different from
zero, locally the remaining singularities are conifold singularities.

\newsec{More on moduli spaces}

The analysis of the previous section treated the generic branch of
moduli space, but there can also be special branches.  The discussion
depends very much on how many $\zeta\ne 0$ so we discuss each case
separately.  In addition we consider the moduli spaces for $M<n$ which
are also physically relevant.  Finally, we will also be interested in
the Coulomb branch of moduli space.  This is defined in the $p$-brane
theories with $p<3$ obtained by naive dimensional reduction as follows:
this theory contains scalar partners of the gauge field, which
we can write as real adjoint fields $X^i$.  Their
potential will be minimized by any vevs with $[X^i,X^j]=0$ and if $\zeta=0$
and $\phi=0$ this is generically
a supersymmetric vacuum with unbroken $U(1)^n$.
The Coulomb branch of interest to us is then the moduli space of
such vacua with general $\phi_i$.

\subsec {The undeformed moduli space}

In the orbifold limit, when we set $\xi _i$ to zero, the equation that
describes the generic branch is

\eqn\nonom{M_{11\dots 1} M_{22\dots 2} M_{33\dots 3}=(-1)^{n-1}M_{123}^n}

This moduli space has three fixed (complex) lines of singularities, all three
with $M_{123}=0$: $M_{11..1}=M_{22..2}=0$, $M_{22..2}=M_{33..3}=0$ and
$M_{11..1}=M_{33..3}=0$. These three fixed lines correspond to the three
lines of $\BC ^2/\BZ_n$ singularities of $\BC ^3/\BZ_n\times \BZ_n$.

The fixed lines also are in correspondence with the moduli space
for $M=1$. For $M=1$ the F-flatness conditions become
$\phi_1\phi_2=\xi_3$ (and permutations) and for $\xi_i=0$ the solutions
are just the fixed lines.  This also generalizes to $M>1$: the Coulomb
branch allows a single non-zero $\phi_i$ in each of the $U(1)$ factors
and $i$ can be different in each factor.

\subsec {Turning on one modulus}

Taking $\xi _3\neq 0$, the moduli space is now given by
\eqn\unnom{M_{11..1}M_{22..2}M_{33..3}-M_{33..3}\xi _3^n=(-1)^{n-1}M_{123}^n}

In this moduli space, two of the original fixed lines, given by $M_{11..1}
M_{22..2}=0$, (plus $M_{33..3}=M_{123}=0$) are deformed to a smooth fixed
$\BC^*$ at $M_{11..1}M_{22..2}=\xi _3^n$, $M_{33..3}=M_{123}=0$.
This fixed line is the only singularity and again it
corresponds to the moduli space
for $M=1$: $\phi _1\phi _2=\xi _3$ and $\phi _3=0$.

To ease the notation, let's rewrite \unnom\ as $xyz-z\xi_3 ^n=t^n$. For a
fixed value of $z\neq 0$, this can be written as $xy={\tilde t}^n+\xi_3^n$,
which is a resolution of an $A_{n-1}$ singularity.  This resolution can
be blown up introducing $n-1$ independent homology 2-cycles and a large
set of supersymmetric $\BP^1$'s.
All this is fibered over $z$ and degenerates at the fixed line,
a picture which suggests that on the total space
each of these $\BP^1$'s corresponds to a 3-cycle.

\subsec {$\xi _2$, $\xi _3\neq 0$}

There are no singularities in the $M=n$ moduli space. This matches with the
fact that there are no solutions for the $M=1$ case.
Recall that the analysis is local: what
has happened is that the singularities have been sent to infinity.

\subsec {$\xi _i\neq 0$}

When all the moduli are turned on, there are $n-1$ singularities on the
$M=n$ moduli space. On the other hand, there are two solutions for $M=1$:
$\phi _i = \pm(\xi _j \xi _k/\xi _i)^{1/2}$ (with the same sign taken for
all $i$). In this case, the correspondence we have observed so far between
singularities of the $M=n$ moduli space and solutions of $M=1$ is no longer
evident.

In the appendix we argue that for the regular representation there is only a
finite number of solutions with $M_{\underbrace {ii..i}_k}\neq 0$, for $k<n$.
We can display $n+1$ of them (actually $M+1$ of them, for $M\leq n$)

\eqn\isol{
\phi _i =+\left ({\xi _j \xi  _k\over \xi _i}\right )^{1/2}
\hbox {I}_{M-p,p}\;\;\;\hbox {I}_{M-p,p}\hbox{=diag}
(\underbrace{+,+..+}_{M-p},\underbrace {-,-,..-}_p).
}

These $n+1$ special solutions (the only commuting ones) seem to correspond
to the different possibilities of distributing $n$ objects between the
two solutions of $M=1$. The gauge invariant polynomials for the special
solutions are

\eqn\isol {M_{ii..i}={n-p+(-1)^np\over n}\left ({\xi _j \xi  _k\over \xi _i}
\right )^{n/2}\;\;\;M_{123}={n-2p\over n}\left ({\xi _1 \xi  _2\xi _3}
\right )^{1/2}}

and for $M=n$ generically don't coincide with the singularities.

\newsec{Fractional branes}

We now discuss the interpretation of the Coulomb
branch of the gauge theory.  For $p$-branes with $p<3$ the world-volume
theory is the naive dimensional reduction of the $d=4$ theory we
described and contains scalar partners of the gauge field, which
we write as the real adjoint fields $X^i$.  Their
potential will be minimized by any vevs with $[X^i,X^j]=0$ and if $\zeta=0$
and $\phi=0$ this is a supersymmetric vacuum with unbroken $U(1)^n$.

Not only does this branch have $n$ moduli describing positions in the
space transverse to the orbifold but it is clear from world-sheet
considerations that it looks like a gravitational and RR source of strength
$1/n$ (compared to the original $p$-brane) at each of these points.
Thus the Coulomb branch describes fractional branes
bound to the fixed point, just as for the case
without discrete torsion \frac.
In that case these were interpreted as branes
wrapped around hidden two-cycles.
However, the singularities with discrete torsion do not contain two-cycles
in the usual string theory sense
(we did not see the corresponding closed string states)
so the story must be rather different here.

The most obvious difference is that, since there is a unique irreducible
projective representation, the fractional branes are classified
by a single conserved $\BZ_n$ quantum number.  This shows
up in the fact that the gauge theory is labelled by the single integer $M$,
and if $M\ge n$ we can find mixed Coulomb-Higgs branches embedding the Higgs
branch in any $U(n)$ subgroup of $U(M)$.  Thus $n$
fractional branes can annihilate to form a conventional brane.

The appearance of fractional branes in these models raises questions
concerning their interpretation, their location and their charge.
Let us raise these issues in turn.

\subsec{Interpretation}

In previous cases, fractional
branes were interpreted as wrapped $p+2$ branes.
For the unresolved orbifold and when one fixed line is resolved,
this picture makes some sense here.  Since the fixed lines are fixed
under a single group element, locally the geometry is $\BC^2/\BZ_n \times \BC$
and we expect to see branes wrapped around these hidden two-cycles.
After the resolution, one can check that the volume (integral of the
holomorphic
two-form) of the two-cycles increases with distance from the origin
(in the third coordinate), which explains why the branes will
then be confined to the neighborhood of the origin.

However these two-cycles are not homology two-cycles so it is not
obvious why such wrapped branes should be stable.
Indeed there is only a conserved $\BZ_n$ quantum number.
This strongly suggests that the objects have some interpretation as branes
wrapped about torsion 2-cycles, which could make contact with the proposal
of \stable.

This proposal had two parts.  The first point was that for geometric
compactifications, discrete torsion is naturally identified with $\int B$
over the torsion part of the 2-cohomology, or more precisely with
${\rm Hom}(H^2(M),\BC^*)$.  This is in agreement with our suggested
interpretation, as the $\int B\wedge C^{(1)}$ on the D$2$-brane would lead to
the correct fractional D$0$-charge.

The second part of their proposal was that orbifolds with discrete torsion
would admit natural resolutions (possibly not Calabi-Yau) which exhibited
this torsion explicitly.  We have nothing to say about this beyond the
comment that if so, it should also be true for our resolved moduli spaces.

In light of the connection between D-brane charges and K-theory \dak,
we can contemplate the possibility that the presence of these torsion
2-cycles can be confirmed by studying the K-theory of the moduli space,
rather than its cohomology (anyway, recall that we are being a little
bit cavalier with the use of the word ``cohomology'' for our varieties).

As was pointed out in \ooguri, independently of
the details of the variety, in the presence of discrete torsion the available
susy cycles are three-cycles, but we don't have $D(p+3)$-branes to wrap about
these cycles!
Take \IIa\ for definiteness;
the natural objects analogous to the fractional branes of \frac\ in
this situation would be strings
coming from wrapped D$4$-branes.  The three-cycles are not localized to
the fixed points but instead are the two-cycles of the resolved
$\BC^2/\BZ_n$ fixed lines, suspended over a real line connecting two
conifold singularities.

It is not surprising that we did not see these objects
in the framework discussed here but we are left with the interesting
question: can we develop a gauge theory (or other construction) of
these strings ?  Perhaps a D$2$-brane with one dimension
stretched along a fixed line could be deformed into two of these objects.

\subsec{Location of fractional branes}

Recall that the moduli space for $M=1$ with all $\xi$ non-zero consisted
of two points
$\phi^{(\pm)}_i = \pm(\xi _j \xi _k/\xi _i)^{1/2}$.
Thus there are two physically (though not topologically)
distinct elementary fractional branes.

One can compute the values of the invariant polynomials for these
configurations and intuitively, one might have expected these
to be identifiable with singularities of the moduli space for the
regular representation.  Since there are $n-1$ singularities and two
fractional branes, this intuition already has problems.
In fact the invariants corresponding to these
points are not on the generic branch at all!

This situation persists for combinations of fractional branes.
Let us consider $n$ fractional branes, for which there are $n+1$
distinct combinations of the two solutions $\phi^{(\pm)}$.
Only for $n$ even and equal numbers $N_+$ and $N_-$
of $\phi^{(+)}$ and $\phi^{(-)}$
is this point on the generic branch (and then it is a singularity).

Thus the decay of the $n$ fractional branes to a regular brane almost
always encounters a potential barrier.  Another way to see this is to
compute the value of the superpotential $W(\phi)$ at the various solutions.
If we consider domain wall solutions of the four dimensional gauge theory
interpolating between the two solutions
(or reduce to two dimensions), they will be BPS with central charge equal
to the difference $W(\phi_a) - W(\phi_b)$.
It is easy to check that the Higgs branch has $W=0$, while the fractional
brane solutions have $W=\sqrt{\zeta_1\zeta_2\zeta_3}(N_+-N_-)$.

Another interesting definition of the `location' of the fractional branes
would be to consider a larger theory containing both a
regular and a fractional brane, and find the values of the regular
brane moduli for which additional massless states appear in the
system.  We leave this as a problem for future research.

The picture looks very non-geometric, on any scale comparable with $\zeta$,
not just at the singularity.
This leads to potential paradoxes in the large $R_{11}$ limit as we can
take $\sqrt{\zeta} >> l_{p11} = g_s^{1/3} l_s$, and at these scales
we expect that M theory is geometric.
It is not too clear if these are actual paradoxes.
For one thing, we did not compute metric data and it might be that
the $O(\sqrt{\zeta})$ separation in complex coordinates does not translate
into an $O(\sqrt{\zeta})$ distance.
Even if it does, if we try
to take the large $R_{11}$ limit
a la Matrix theory, the large $N$ limit might produce
a very different picture.
It would be interesting to find some geometric picture which can evolve
smoothly to the one presented here.

\subsec{Charge quantization}

Fractional branes carry $1/n$ of the charge
of the original branes. At first sight this seems to conflict wih the Dirac
quantization condition for D-brane charges. In models without discrete torsion,
a way which has been proposed out of this paradox is by analogy with the answer
to the familiar question of why the fractional charge of quarks is not
inconsistent with the Dirac quantization condition for monopoles \corri.
In both cases the fractionally charged objects carry an additional charge, and
that can modify the Dirac quantization condition. In the case of quarks
this charge is of course color, and if the group is $U(1)_{EM}\times SU(N)_c$
the quantization condition is $eg=2\pi n/N$. For fractional branes, both in
the ALE case and in the $T^6/\BZ_n \times \BZ_n$ without discrete torsion,
from $h^{1,1}\neq 0$ we see that there are 2-cycles, and the original 3-form is
reduced on them to a 1-form. Fractional branes are charged under this reduced
1-form \frac, and that could solve the paradox.

However, in the present case, this cannot be the resolution, as
there are no RR gauge fields in twisted sectors for the objects to
couple to.  Consider D$0$'s in \IIa\ for definiteness; we found that all
the twist states corresponded to elements of $H^{1,2}$,
whose RR partners are scalars.

Checking the Dirac condition requires introducing the D$6$-brane
and either computing properly normalized
charges for the two objects or else computing their interactions directly
(perhaps from the annulus diagram, along the lines of \pol).  The first
computation involves the volume of the internal space and when passing
to the orbifold, gives a factor of the order of the group which could
resolve the problem.  However, an indication that something more unusual
is going on here is that the order of the group is $n^2$, compared with
the $1/n$ we are trying to account for.

The second computation can be done easily if we take the open string
(gauge theory) point of view and restrict to the massless sector.
In \berkdoug\ it was shown (in a very similar matrix theory problem)
that the magnetic monopole interaction between a D$p$ and D$(6-p)$ brane
can be derived from the Berry phase of the Hamiltonian describing
fermionic strings stretched between these objects.
These fermions are a doublet of the $SO(3)$ transverse to both branes
and their Hamiltonian is simply $H =\vec\sigma\cdot \vec X$; it is well
known that the Berry phase for this Hamiltonian is described by the
magnetic monopole gauge potential.  On the other hand, massive string modes
will always come in pairs with cancelling Berry phase.  In particular we
can ignore the winding states around $T^6$ for this computation.

Thus the interaction relevant for the Dirac quantization condition can
be computed simply by counting fermionic strings.  It also
means that perturbative consistency of string theory
guarantees that D-branes will satisfy the Dirac rule.

The matrix $N_{ab} = {\rm Tr}_{ab}\ (-1)^F$ counting massless Ramond doublets
(with chirality)
between the $a$'th and $b$'th brane is the
CFT analog of the intersection form in a geometric compactification.
It is an index and thus will not vary under continuous deformations.
In the geometric case, if we consider
a set of branes wrapping an integral homology basis,
this form must
be integral and unimodular by Poincar\'e duality, proving that the
Dirac condition is saturated.
By computing $N_{ab}$ in a CFT compactification
we can check that a particular set of D-branes also saturates the
Dirac condition and is thus complete.

D$6$-branes wrapped about the internal space
in these orbifold theories are defined by the same projection
\projcon\ of flat space gauge theory but now with the space-time operation
$R$ acting on the world-volume coordinates as well.  If we start with an
$N$-dimensional representation $\gamma$ this will lead to a $U(N)$ gauge
theory on the quotient space $T^6/\Gamma$ with specific boundary conditions.
depending on the choice of representation.

The 0--6 strings in our problem will then give fermionic fields $\chi$
satisfying the projection
\eqn\zerosix{
(\gamma^{(6)}(g))^{-1} \chi \gamma^{(0)}(g) = R(g) \chi .
}
In fact $R(g)=1$ here (in contrast to the vertex operators $V(\zeta)$
we discussed earlier which do transform under $U(1)^3$ rotations).
This is clear because the Ramond states must form a representation of
the $SO(6)$ rotating the DN directions before applying the orbifolding;
as is well known this is a singlet representation.

Let us briefly discuss the case without discrete torsion first
(we will return to this elsewhere).
We first note that
in contrast to the $0$-branes it is more reasonable to call the object
with $N=1$ the elementary $6$-brane since its world-volume theory is $U(1)$
gauge theory on the quotient space.  Both it and the fractional $0$-brane
will carry one-dimensional representations of $\Gamma$ and the projection
\zerosix\ implies that only if the two branes carry the same representation
$\gamma$ will a 0-6 string survive: we have $N_{ai} = \delta_{a,i}$ in
this basis.  Thus the Dirac condition is saturated if we include the
$N=1$ $6$-branes.

For the case with discrete torsion, the minimal theory then takes
$\gamma^{(0)}=\gamma^{(6)}=\gamma_1$, an $n\times n$ representation, to
describe a fractional $0$-brane and an elementary $6$-brane.
A single component $\chi\propto {\bf 1}$ survives the projection and
we conclude that these two objects satisfy the Dirac condition with
the minimal flux quantum.

However, the striking feature of this elementary $6$-brane is that
it is a $U(n)$ gauge theory on the quotient manifold, not a $U(1)$
theory, since $\gamma _1$ is an $n$-dimensional representation.
As we commented above,
the equations \projcon\ have a solution for generic $U(n)$ gauge
fields in one fundamental region $T^6/\Gamma$; they simply determine
the corresponding fields in the other fundamental regions.

Thus the final resolution of our paradox turns out to be that, in the
case with discrete torsion, the elementary $6$-brane is actually an
``$n$-fold bound state,'' in the sense that it carries $U(n)$ gauge fields.

\newsec{Conclusions and further questions}

We derived D-brane gauge theories for $\BC^3/\BZ_n\times\BZ_n$ orbifolds
with discrete torsion and studied the moduli space of a D-brane at a
point (say a D$0$ in \IIa).  We were able to find a simple exact equation
for this moduli space as a subvariety of $\BC^4$.
In agreement with expectations the
closed string moduli deform the moduli space and resolve the fixed lines
but do not allow fully resolving the singularity.

However the detailed results for $n>2$ conflict with the intuition that
discrete torsion, being non-geometrical,
had to be ``located'' at a particular singularity.
Instead we find $n-1$ conifold singularities separated (in complex
parameter space) by $O(\sqrt{\zeta})$, where $\zeta$ is the scale of
the resolution parameter.

Was there any real basis for the intuition that discrete torsion would
be concentrated at one singularity?
Probably not.  One argument against this is that, in the global
context (consider $T^6/\BZ_3\times\BZ_3$ for example), the value of
discrete torsion is not independently adjustable at each of the singularities
but rather is a global choice to be made for the entire orbifold.
{}From this point of view any number of singularities might have appeared
in the resolution with discrete torsion being a global invariant not
necessarily detectable at any one of them.

We also found fractional branes which are BPS and carry a conserved
$\BZ_n$ quantum number.  Some of their properties are consistent with
the idea that these are $p+2$-branes wrapped about a zero volume
torsion 2-cycle of the type suggested in \stable.  Other properties -- there
are two elementary fractional branes distinguished by a non-topological
quantum number, and the annihilation of $n$ branes to a regular brane
encounters a potential barrier -- are rather peculiar.

The $T^6/\BZ_2\times\BZ_2$ model is mirror to the same orbifold without
discrete torsion.  Since this can be completely resolved,
this should provide a geometrical picture of the
fractional branes in terms of $3$-branes.  Presumably the potential
barrier would mean that
a supersymmetric $T^3$ cycle (mirror to the D$0$) is homologous to a sum
of $n$ supersymmetric cycles without moduli (say $S^3$'s) but that they
are connected only by deforming through non-supersymmetric cycles.

The fractional branes carry charge
$1/n$ yet satisfy the Dirac quantization condition; related to this,
the elementary conjugate $6-p$-brane is a ``bound state'' in the sense
that its world-volume theory is a $U(n)$ gauge theory.

It will be quite interesting to
make contact between these results and the K-theory description
of D-brane charge, and to develop either a geometrical interpretation of
discrete torsion or perhaps a noncommutative geometric description.

\medskip

{\bf Acknowledgements}

B.F. would like to thank Duiliu-Emanuel Diaconescu, Jaume Gomis, and
Christian Roemelsberger for useful discussions and encouragement.
M.R.D. would like to thank Edward Witten for a discussion.

{\appendix A { }}

In this appendix we prove that on the moduli space of our $U(n)$
gauge theory, when all $\zeta\ne 0$, any gauge invariant polynomial $I$
can be written as a polynomial in terms of $M_{ii..i}$ and $M_{123}$:
$I=P_I(M_{11\ldots 1},M_{22\ldots 2},M_{33\ldots 3},M_{123};\zeta_i)$.
This means that the moduli space is a subvariety of $\BC^4$
defined by polynomial equations,
which could be obtained in principle by finding a complete set of
relations between gauge invariants and substituting these linear relations.

The simplest relation comes from
taking the trace of the F-flatness conditions \dfflat: $M_{ij}=\xi_k$.

The idea is, given some invariant $M_{i_1i_2\ldots}$, to use the
deformed F-flatness conditions \dfflat\ and the cyclic property of the trace,
to shift around one of the indices, until the original ordering is recovered.
If this comes with a phase, we can express the invariant in terms of
lower degree invariants. The simplest example will
illustrate the general method.
$$\eqalign {& \phi _1\left (\phi _1\phi _2-\epsilon ^{-1}\phi _2\phi _1
\right )=-\zeta _3\phi _1\Rightarrow M_{112}-\epsilon ^{-1}M_{121}=-
\zeta _3M_1\Rightarrow\cr
& \left (1-\epsilon ^{-1}\right ) M_{112}=-\zeta _3M_1
\Rightarrow M_{112}=M_1M_{12}}.$$
Another example (which we leave as an
exercise for the idle reader) is that $M_{iijj}=M_{ijij}= M_{ij}^2$.

However, this method does not work for every invariant. For invariants
with all indices equal, as $M_{11..1}$, the method does not even apply.
This is not the only case for which the method fails to express an
invariant in term of lower degree invariants: if $M_{ij..k}$
has charge $(n_1,n_2,n_3)$,
then when $n_1-n_2 \equiv n_2-n_3\equiv n_3-n_1\equiv 0\;(n)$ --
i.e. $n_1\equiv n_2\equiv n_3\;(n)$ -- we obtain instead
an identity involving only
lower degree invariants. For example, for $n>2$
$$\eqalign {& M_{112233}=e^{-{2\pi i\over n}2}M_{122133}-\left (1+
\epsilon ^{-1}\right)\zeta _3M_{1233}=\cr &=M_{112233}+\left (
1+\epsilon ^{-1}\right)\left (\zeta _2M_{1223}-\zeta
_3M_{1233}\right ) \cr
&\Rightarrow \xi _2M_{1223}=\xi _3M_{1233}}$$
so using this method we fail to express $M_{112233}$ in terms of lower degree
polynomials, but instead obtain a relation between lower degree polynomials.

At this point our list of potentially independent polynomials on the moduli
space consists of all the polynomials of the form $M_{ii.i}$, and the
polynomials $M_{11..122..233..3}$ with $n_1\equiv n_2 \equiv n_3 (n)$.
To obtain more relations we assume that our matrices are $n$ dimensional
and use the Cayley-Hamilton theorem.
This states that a matrix $\phi$ will solve the equation
\eqn\cayley{
P_\phi(\phi)=0
}
where $P_\phi(x)$ is the characteristic polynomial of $\phi$, a
polynomial of degree $n$ whose coefficients are polynomial in $\Tr \phi^k$
for $k\le n$.  The simplest proof follows from the fact that diagonalizable
matrices are a dense subset.

Since \cayley\ is a matrix relation we have
\eqn\truc {\hbox {tr }A\!P_\phi(\phi)=0}
for any matrix polynomial $A$.
This shows that any invariant of degree greater
than $n$ with $n$ repeated indices can be written in terms of lower
degree invariants.
This reduces our list of potentially independent invariants to
$M_{ii..i}$ with up to $n$ indices plus
$M_{123123..123}$ with strictly less than $3n$ indices.

We can do even better. Now we are going to argue that the situation is the
following: for generic values of $M_{ii..i}$, when the relation
$\xi _1^n M_{\underbrace{11..1}_n}=\xi _2^n M_{\underbrace {22..2}_n}=\xi _3^n
M_{\underbrace {33..3}_n}$ is not satisfied, we can prove that
$M_{\underbrace {ii..i}_k}=0$ for $k<n$. On the other hand, in the special
case when that relation is satisfied, $M_{ii..i}$ for $k<n$ need not to be
zero, (and indeed it is not zero for some solutions), but we nonetheless
argue that $M_{ii..i}$ can only take a finite number of values, so they cannot
parametrize a continuous direction.

To start with, one can very easily prove that

$$\eqalign {M_{\underbrace {11..1}_a \underbrace {22..2}_b
\underbrace {33..3}_c}(e^{{2\pi i\over n}b}-e^{{2\pi i\over n}c})= \cr
[\xi _3 (e^{{2\pi i\over n}b}-1)M_{\underbrace {11..1}_{a-1}
\underbrace {22..2}_{b-1} \underbrace {33..3}_c}-\xi _2 (e^{{2\pi
i\over n}c}-1) M_{\underbrace {11..1}_{a-1}
\underbrace {22..2}_{b} \underbrace {33..3}_{c-1}}]}.$$

Applying this first to $M_{\underbrace {11..1}_{k} \underbrace {22..2}_{k}
\underbrace {33..3}_k}$ and then succesively to the resulting relations we
obtain that for $k<n$

\eqn\mii{\xi _1^k M_{\underbrace{11..1}_k}=\xi _2^k
M_{\underbrace {22..2}_k}=\xi _3^k M_{\underbrace {33..3}_k}}

On the other hand, take $\phi $ in \truc\ to be $\phi _1$ and $A$ in to be
successively $\phi _2\xi _1^{n-1}$, $\phi _2^2 \xi _1^{n-2}$,...
$\phi _2^{n-1}\xi _1$, and in each case then substract the same relation
with $1\leftrightarrow 2$. Using \mii , we obtain

\eqn\nindex{M_{\underbrace {11..1}_n}M_{\underbrace {22..2}_k}\xi _1^{n-k}=
M_{\underbrace {22..2}_n}M_{\underbrace {11..1}_k}\xi _2^{n-k}}

Now if any of the $M_{11..1}$ with less than $n$ indices is not zero,
we can use the last relation (and two similar ones, replacing
$1\rightarrow 3$ in the first, and $2\rightarrow 3$ in the second) to
prove that $\xi _1^n M_{\underbrace{11..1}_n}=\xi _2^n
M_{\underbrace {22..2}_n}=\xi _3^n M_{\underbrace {33..3}_n} $. This shows
that for generic values of $M_{\underbrace {ii..i}_n}$ and $\xi _i$, (that is,
 when the previous
relation is not satisfied), we have $M_{\underbrace {ii..i}_k}=0$ for
$k<n$. What can we say when $\xi _1^n M_{\underbrace{11..1}_n}=\xi _2^n
M_{\underbrace {22..2}_n}=\xi _3^n M_{\underbrace {33..3}_n}$? Plugging this
relation and \mii\ into the characteristic polynomials of $\phi _1$,
$\phi _2$ and $\phi _3$ we learn that in this case $\phi _2=(\zeta _1/
\zeta _2)g \phi_1 g^{-1}$ and $\phi _3=(\zeta _1/\zeta _3)\tilde {g} \phi_1
\tilde {g}^{-1}$. Anyway, we can argue that there is only a finite number
of values the $M_{\underbrace {ii..i}_k}$ can take. To see this take $\phi$
in \truc\ to be $\phi _1$ and $A$ to be succesively $\phi _2$, $\phi _2^2$..
$\phi _2^{n-1}$. Using \mii\ we obtain $n-1$ equations for $M_{ii..i}$, and
this system of equations has a finite number of solutions.

Finally, we have to deal with the polynomials of the kind $M_{123..123}$. We
can not apply \truc\ as it stands, since we have argued that for generic
values of $\xi _i$, $M_{ii..i}=0$ for $k<n$, so the Cayley-Hamilton theorem
reduces to $\phi _i^n-M_{11..1}I=0$. If we consider the characteristic
polynomial of $\phi _1\phi _2$ it does not work neither. If we consider the
characteristic polynomial of $\phi _1\phi _2\phi _3$ and take $A=\phi _1,
\phi _1 ^2...$, we obtain relations for these polynomials. FOr instance, for
$n=3$ we obtain

\eqn\udt{M_{123123}=M_{123}^2+2\epsilon (1-\epsilon) M_{12}M_{31}M_{23}}

So for generic values of $M_{ij}$ we see that $M_{123..123}$ is also determined
in terms of lower degree polynomials. This concludes our argument that on the
moduli space, all the non-zero gauge invariant polynomials can be written in
terms of 4 polynomials: $M_{\underbrace {11..1}_n}$,
$M_{\underbrace {22..2}_n}$, $M_{\underbrace {33..3}_n}$ and $M_{123}$

\listrefs
\end

The authors of \ref \aspi {P. Aspinwall, Morrison "On stable singularities in
string theory"} argue that the situation is the following. We start with a
target space $X^{\#}$. Without discrete torsion in the CFT, we can blow it up
to a smooth target space $Y$. When we add discrete torsion to the CFT, we can
instead deform $X^{\#}$ to a manifold $X$ that still has singularities.
Finally, $X$ can be blown-up by an irrelevant operator to a smooth target
space $\tilde X$, but this $\tilde X$ is not a CY. They conjecture that
$Tors\; \left (H_2(\tilde X;\BZ)\right )=H_2(\Gamma)$.

A reason to be suspicious of the role of additional RR charges even in the
other cases is that the additional charges arise geometrically as a result
of non-zero $\int B$ at the orbifold point; this is a modulus and if
we take it to zero we lose these charges.  Although the fractional branes
become massless at this point and an approach in pure CFT terms breaks down,
the Dirac condition must still be satisfied, now between the pure $0$-brane and
$6$-brane.